\documentclass{sig-alternate-05-2015}
\usepackage{tikz}
\usepackage{array}
\newcolumntype{M}[1]{>{\centering\arraybackslash}m{#1}}
\graphicspath{{./img/}}

\begin{document}

\usetikzlibrary{shapes,arrows,positioning,decorations.pathreplacing, mindmap}
\tikzstyle{block} = [rectangle, draw, fill=blue!20, 
    text width=5em, text centered, rounded corners, minimum height=4em]
\tikzstyle{block1} = [rectangle, draw, fill=red!20, 
    text width=9em, text centered, rounded corners, minimum height=4em]
\tikzstyle{block2} = [rectangle, fill=white!20, 
    text width=5em, text centered, rounded corners, minimum height=4em]
\tikzstyle{line} = [draw, -latex']

\setcopyright{acmcopyright}





%

\title{Multimodal Content Representation and Similarity Ranking of Movies}
%
%
%
%
%

\numberofauthors{2} 
%
\author{
%
%
\alignauthor
Konstantinos Bougiatiotis\\
      \affaddr{Software and Knowledge Engineering Lab,}\\
      \affaddr{Institute of Informatics and Telecommunications,}\\
      \affaddr{National Center of Scientific Research Demokritos, Greece,}\\
      \email{bogas.ko@gmail.com}
2nd. author
\alignauthor
Theodoros Giannakopoulos\\
      \affaddr{Computational Intelligence Lab}\\
      \affaddr{Institute of Informatics and Telecommunications,}\\
      \affaddr{National Center of Scientific Research Demokritos, Greece,}\\
      \email{tyiannak@gmail.com}
}
\date{29 January 2016}

\maketitle
\begin{abstract}
In this paper we examine the existence of correlation between movie similarity and low level features from respective movie content. In particular, we demonstrate the extraction of multi-modal representation models of movies based on subtitles, audio and metadata mining. We emphasize our research in topic modeling of movies based on their subtitles.
In order to demonstrate the proposed content representation approach, we have built a small dataset of 160 widely known movies. We assert movie similarities, as propagated by the singular modalities and fusion models, in the form of recommendation rankings. We showcase a novel topic model browser for movies that allows for exploration of the different aspects of similarities between movies and an information retrieval system for movie similarity based on multi-modal content.

\end{abstract}

%
%

\ccsdesc[500]{Information systems~Multimedia and multimodal retrieval}
\ccsdesc[300]{Information systems~Multimedia information systems}
\ccsdesc[100]{Computing methodologies~Machine learning}

%
%

%
%
\printccsdesc


\keywords{Topic Modeling; Latent Dirichlet Allocation; Movie Audio Analysis; Multimodal Fusion; Information Retrieval }

\section{Introduction}

In order to cope with the overwhelming amount of data, we are in dire need of recommendation systems, to browse through item collections. This is also the case when looking at \emph{motion pictures} in particular. There exist many systems providing movie recommendation services, most of which can be classified into either \textit{collaborative filtering} systems, such as \emph{MovieLens}\footnote{https://movielens.org/}, either \textit{content-based} systems, like \emph{jinni}\footnote{http://www.jinni.com/}, or \textit{hybrid} systems, as is \emph{IMDB}\footnote{http://www.imdb.com/}. However, all these systems rely on human-generated information in order to create a corresponding representation and assess movie to movie similarity, not taking into account the raw content of the movie itself but solely building upon annotations made by the users.

This paper introduces the more ambitious objective of representing each movie \emph{directly from its content}. We are striving to find correlations between low level content similarity from different movie information modules and high level association of those movies. This will lead us to innovative ways of defining movie similarity, explore \emph{latent} semantic knowledge from topic models and boost traditional information retrieval systems with information from heterogeneous content sources. The usage of the multimedia signal of the movies has been limited to particular applications such as emotion extraction\cite{malandrakis:emot} or violent content detection\cite{giannak:viol, nam:viol}. An interesting application where topic modeling is used in the movie domain, aims at creating movie summaries containing those movie scenes that best embody the gist of the topic the movie mainly belongs\cite{boll:ldamovie} to.

The rest of this paper is organized as follows. Firstly, the general workflow and details of the proposed method are explained (Section \ref{sec:prop}). We then present our data collection and ground truth generation methodology (Section \ref{sec:data_gt}). In the following section (Section \ref{sec:res}) the experimental results are presented. We close by drawing conclusions and outlining topics for further research (Section \ref{sec:fut}).

\section{Proposed Method}\label{sec:prop}
\subsection{General Workflow}
\begin{figure}[ht]
    \centering
    \includegraphics[width = 0.5\textwidth]{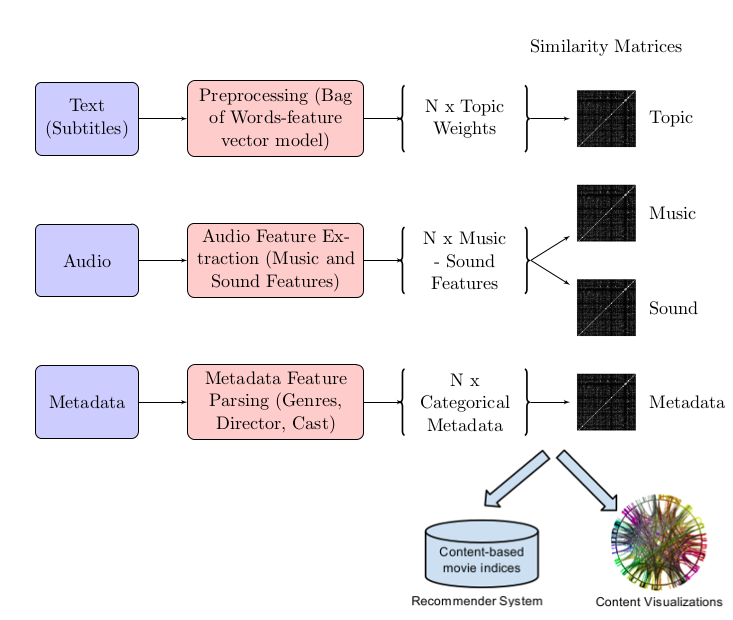}
    \caption{Workflow diagram of the proposed method}
    \label{fig:workflow}
\end{figure}

The overall scheme of the methodology described in the current work is presented in  Figure \ref{fig:workflow}. In summary, the following steps, with regard to the different modalities, are carried out :
\begin{itemize}
\item Text Analysis: Preprocessing, followed by the training of a topic model (through \emph{Latent Dirichlet Allocation}), of the subtitles for each movie, in order to represent the corresponding movies as vectors with topic weights.
\item Audio Analysis: two supervised audio classification models that result in a music-related and an audio event-related representation.
\item Metadata Analysis: Parsing metadata information about each movies' cast, director and genre into categorical feature vectors.
\item Data Fusion and Content Similarity: Fusing the similarity matrices that were generated through the previous steps, we yield recommendations for each movie.
\end{itemize}

\subsection{Subtitles Analysis}

We start by applying essential preprocessing steps on each subtitles' document. The main steps involved in this preprocessing stage are:
(a) Regular expressions removal (filtering out timestamps and non-textual characters), (b) Tokenization-case folding (tokenizing and reducing letters to lower case), (c) Lemmatization based on \emph{WordNet} database\cite{fellbaum:wordnet} (unifying variations of the same term due to inflectional morphology)  

As a second preprocessing stage, we apply \emph{word filtering}. Firstly, common and movie-domain specific stopwords are removed. Then, we remove words which provide low information for each document. These are words with low intra-document and high inter-document frequency. At the end of this preprocessing set of procedures, we acquire the \textit{bag of words}(\textit{BoW}) representation for the documents, which further leads to a multi-dimensional vector of term frequencies for each movie.

BoW representation is far too sparse and highly-dimensional to be used efficiently, so we deploy a topic modeling algorithm, namely \emph{Latent Dirichlet Allocation}(\emph{LDA})\cite{blei:lda}, a \emph{probabilistic topic model} where the fundamental idea is that all the documents(movies) in the collection share the same set of topics, but each document exhibits those topics in different proportions. We used a \emph{Collapsed Gibbs Sampling} version of the algorithm\cite{mccallum:mallet} and after numerous empirical evaluations we settled for $T=55$ topics as the optimal value.

\begin{figure}[ht]
\centering
\includegraphics[width = 0.4\textwidth]{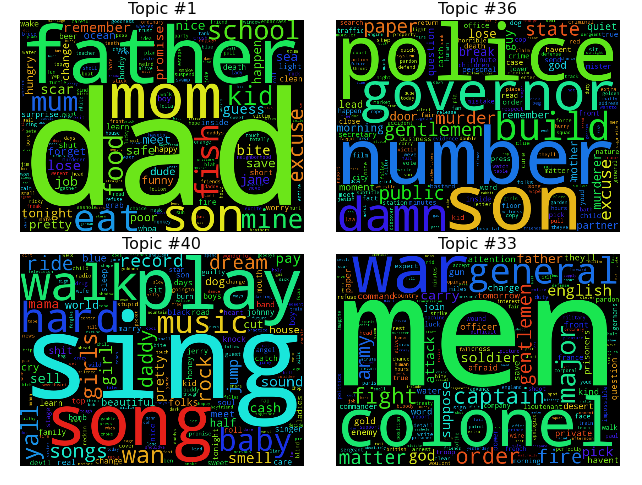}
\caption{Word Clouds examples for 4 Topics}
\label{fig:clouds}
\end{figure}

In\textbf{ Figure \ref{fig:clouds}} we depict $4$ topics generated from our movie collection, presented as word clouds, where the size of each word is proportional to the importance of the word for this topic. If we observe the resulting topics, we can see that they are well formulated and coherent. For example, the top left topic is highlighted by words such as \textit{dad, father, mom, son, school}, defining a family related topic while the bottom right exhibits mainly words \textit{like men, colonel, war, general}, defining a war related topic.

Let it be noted that  \textit{tf}-\textit{idf} weighting\cite{salton:information} and \textit{Latent Semantic Indexing}\cite{deerwester:lsi} methods, with $K=35$ latent dimensions, have also been applied, for benchmarking purposes and comparison against the proposed \textit{LDA} model.

\subsection{Audio Analysis}
The audio signal  is a very important channel of information with regards to a movie's content: music tracks, musical  background themes, sound effects, speech, sound events, they all play a vital role in forming the movie "style". Therefore, a content representation approach should also take into account these aspects of information. Towards this end, we have extracted two types of information: (a) music-genre statistics and (b) audio event statistics. 

In particular, we have trained two separate supervised models using Support Vector Machines, in order to classify all movie segments to a set of predefined classes related either to audio events or musical genres. Towards this end, the pyAudioAnalysis library has been used \cite{giannakopoulos2015pyaudioanalysis} to extract audio features both in a short-term and in a long-term basis. Then, each long-term segment (represented by a vector of long-term audio features as described in \cite{giannakopoulos2015pyaudioanalysis}) is fed as input to both musical genre and audio event classifiers. The result of this process is a sequence of music-genres and a sequence of audio events. The final representation that corresponds to the whole movie is provided by to vectors that represent the percentage of each musical-genre or audio event class. 

Note that, in order to train the two classifiers a separate and independent dataset has been annotated. The adopted classes for musical genres are: blues, classical, country, electronic, jazz, rap, reggae and rock. The audio event classes are: music, speech, three types of environmental sounds (low energy background noise, abrupt sounds and constant high energy sounds), gunshots, screams and fights.  

\subsection{Metadata Analysis}
Feature extraction from metadata is much more straightforward. Utilizing publicly available information regarding the cast, the directors and the genres of the movies in our collection from \emph{IMDB}, we create a categorical vector for each movie, where each cell contains a binary value, $0$ or $1$, denoting relation between the movie and the categorical value. 

\subsection{Content Similarity and Data Fusion}
Having represented the movies as feature vectors, we can define similarity between these vectors to correspond to the similarity of their respective movies. We compute the \textit{cosine similarity} between all movie pairs ($\vec{m_a}, \vec{m_b}$), in the different representation spaces:

\begin{equation*}
 CosSim(\vec{m_a}, \vec{m_b}) = \frac{\vec{m_a}\times \vec{m_b}}{\|\vec{m_a}\|\times \|\vec{m_b}\|}   
\end{equation*}

This results in a \textit{similarity matrix} between movies for each modality. In order to combine these content-specific similarities we adopted a simple weighting scheme between the similarity matrices, where the optimal weights for each modality are pinpointed after extensive experimentation.

\section{Dataset}\label{sec:data_gt}
\subsection{Data Description}
In order to demonstrate the usefulness of low-level content representation of the movies for similarity purposes, as well as browsing and exploration of content, we have compiled a real-world dataset of 160 movies. These movies have been selected from the \emph{Top 250 Movies}\footnote{http://www.imdb.com/chart/top}. Our purpose was to use movies that are widely known and therefore the quality of the results can be easily assessed. Moreover, the dataset is populated with different types of movies to avoid metadata-specific bias, such as genre or casting. The subtitles were downloaded from an open source database\footnote{http://www.opensubtitles.org/en/search} and were hand-checked for mistakes. 

\subsection{Ground-truth generation}
However, to evaluate the similarity rankings generated by the different modalities we need a \emph{ground-truth} similarity between the movies-documents of the dataset, against which we can pitch our results. Towards this end, we used the \emph{Tag-Genome}\cite{vig:tag} dataset to create a ground-truth similarity matrix between the movies. Every movie is represented as a vector in a tag-space with $\approx1100$ unique tags. The tags can be a wide variety of words-phrases such as adjectives("funny", "dark", "adopted from book"), nouns("plane", "fight") metadata("tarantino", "oscar") and so on, that act as descriptors for the movies. Having this representation for each movie we obtained the ground-truth movie similarity matrix, as before. 

\section{Experimental Results}\label{sec:res}
Moving on, in order to appraise the quality of the similarities of the individual content representation models, we utilized the similarity rankings created by the aforementioned matrices. Specifically, for each movie we are interested in the similarity ranking of the first recommendation generated by each model. We calculate the median position, of the first recommendations over all the movies, as ranked in the ground truth similarity matrix. This information-retrieval measure conveys the similarity ranking accuracy for each model.
Let us note here that, we used the median of the rankings as it is more robust in skewed collections, like these of the rankings for each model.

As a supplementary measure we calculate the percentage of the recommendations that belong to the top 10 recommendations for each movie, according to the ground truth similarity. This serves as a \textit{recall} type of measure, indicating the  sensitivity of each model.

\textbf{Tables \ref{tab:sing}} and  \textbf{\ref{tab:fus}} presents the results for each individual model, as well as, the best data fusion models, over the possible combinations of models. The best combinations arose after lengthy experimentation.

\begin{table}[!htb]
      \centering
        \begin{tabular}{M{3.2cm}|M{2cm}|M{2cm}}\hline
        Model & Median Ranking for 1st Rec & Top 10$\%$ of 1st Rec  \\ \hline
        Tf-idf & 18 & 42.5\\ 
        LSI & 15.5 & 41.8\\
        LDA & 15.5 & 44.3 \\  \hline
        Audio (A) & 51 & 13.1\\ 
        Music (M) & 57 & 14.3\\ \hline
        Metadata (MD) & 8 & 57.5\\ \hline
        \end{tabular}
        \caption{Singular Modalities}\label{tab:sing}
\end{table}

\begin{table}[!htb]
        \centering
        \begin{tabular}{M{3.2cm}|M{2cm}|M{2cm}}\hline
        Model & Median Ranking for 1st Rec & Top 10$\%$ of 1st Rec  \\ \hline
        MD + LSI & 4 & 65.0\\
        MD + T + A & 3 & 65.6 \\  
        MD + LSI + A + M & 3 & 68.7\\ \hline
        \end{tabular}
        \caption{Fusion Models}\label{tab:fus}
\end{table}

It can be clearly seen that as far as individual modalities are concerned (\textbf{Table \ref{tab:sing}}) the metadata model outperforms the rest. That is to be expected, as metadata are high-level features, attributed to movies by humans and, to some extend, correlated with the tag representation of the ground truth similarity. Examining the content-representation models we can see that regarding the textual models, \textit{LDA} mar\-ginal\-ly outperforms \textit{LSI} in the percentage measure, while tying in similarity ranking. On the other hand, the audio modalities as standalone models aren't suitable for recommendations.

However, after inspecting (\textbf{Table \ref{tab:fus}}) we can see that expanding the fusion with more models, enhances the performance of the resulting fusion models. Overall, \emph{we see at least $50\%$ improvement on the results of the best singular model(namely metadata) with the addition of the content models}. This is a really promising outcome, since it proves that low-level movie information can lead to a performance boosting of a content-based recommendation system. 

\section{Conclusion and Future Work}\label{sec:fut}
In this short paper, we adumbrated a multimodal similarity model for movies, based on raw content. In particular, we focused on topic modeling techniques on the subtitles' content. The basic outcomes of our research, as shown above, are the following:
\begin{enumerate}
\item A complete methodology for similarity extraction and retrieval for movies based on low-level features.
\item The most important and promising outcome of the experimentation is that low-level feature models exploit latent information that boost the performance of human-generated information models. They can therefore be adopted in the context of a multimodal content-based recommendation system.
\item Experimentation has shown that \emph{LDA} and \emph{LSI} latent spaces, offer good representations for the movies, with negligible differences in results. \emph{LSI} is by far more efficient in terms of memory, time and complexity to  \emph{LDA}, however \emph{LDA} offers a much more coherent topic mapping of the movies, suitable for topic browsing and similarity discovery. This is portrayed in the qualitative example in \textbf{Figure \ref{fig:topic_movie}}.

\begin{figure}[!ht]
\centering
\includegraphics[width = 0.45\textwidth]{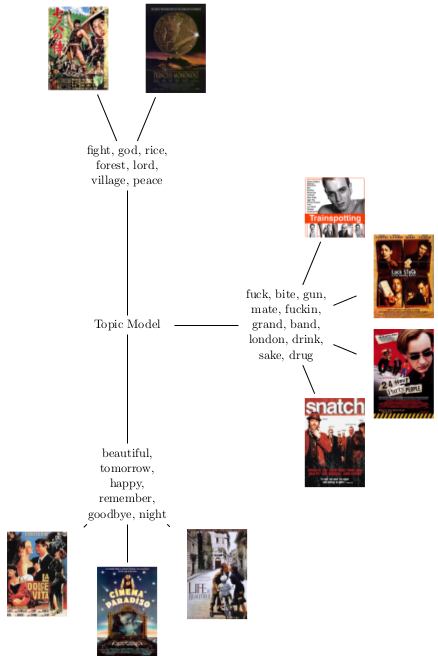}
\caption{Topics-Movies association diagram}
\label{fig:topic_movie}
\end{figure}

We demonstrate how our topic model has clustered certain movies together based on their relevance through specific topics. In the top of the chain we have the most striking example, where \emph{Princess Mononoke} and \emph{Seven Samurai} were grouped together. These movies are not similar according to conventional recommendation systems because one is an animation film and the other is an epic war drama. However, both are set in Japanese rural villages during feudal ages, with striking Japanese cultural elements such as strong religious beliefs, contact with nature and even consumption of rice. All these connecting details are captured in a topic whose main words are \emph{fight}, \emph{god}, \emph{rice}, \emph{forest}, \emph{lord}, \emph{village}, as shown in the figure. Likewise for the rest illustrated examples.

\end{enumerate}

These results verify the core ideas of this work and inspire many future directions for our research. In particular:
\begin{itemize}
\item Implement scalable and efficient methods by adding more movies to our database and testing different topic models such as Hierarchical Dirichlet Processes\cite{teh:hdp}.
\item Experiment with visual features from the movies, augmenting this hybrid fusion system with rich visual ques.
\item Examine more sophisticated fusion schemes and add \textit{user preferences} (collaborative filtering) towards a complete recommendation system.
\end{itemize}

%
\bibliographystyle{abbrv}
\bibliography{bibl}  
%
%
\end{document}